\documentclass[conference]{IEEEtran}
\IEEEoverridecommandlockouts
\usepackage{cite}
\usepackage{amsmath,amssymb,amsfonts}

\usepackage{graphicx}
\usepackage{textcomp}
\usepackage{xcolor}
\usepackage{comment}
\usepackage{tabularx}
\usepackage{booktabs}
\usepackage{caption}
\usepackage{multirow}
\usepackage{subcaption}
\usepackage{todonotes}
\usepackage{algorithm}
\usepackage{algpseudocode}
\usepackage{amsmath}

\usepackage[most]{tcolorbox}

\definecolor{lightgreen}{rgb}{0.894,0.961,0.949}
\definecolor{darkgreen}{rgb}{0.0,0.576,0.533}

\tcbset{
  findingbase/.style={
    enhanced, breakable,
    arc=0mm, boxrule=0mm,
    colback=lightgreen,
    top=2mm, left=2mm, right=2mm, bottom=2mm,
    leftrule=1mm,
    frame hidden, 
    borderline west={6pt}{0pt}{darkgreen}, 
    colbacktitle=darkgreen, 
    coltitle=white,         
    fonttitle=\bfseries     
  }
}

\newtcolorbox{finding}[2][]{%
  findingbase,
  title={#2},
  #1
}

\usepackage{array}
\def\BibTeX{{\rm B\kern-.05em{\sc i\kern-.025em b}\kern-.08em
    T\kern-.1667em\lower.7ex\hbox{E}\kern-.125emX}}

\newcommand*\circled[1]{\tikz[baseline=(char.base)]{
            \node[shape=circle,draw,inner sep=1pt,font=\sffamily\footnotesize] (char) {\textbf{#1}};}}
    
\begin{document}

\raggedbottom


\title{What Makes Software Bugs Escape Testing? Evidence from a Large-Scale Empirical Study\\
}


\author{
\IEEEauthorblockN{
Domenico Cotroneo\IEEEauthorrefmark{1},
Giuseppe De Rosa\IEEEauthorrefmark{2}\IEEEauthorrefmark{3},
Cristina Improta\IEEEauthorrefmark{2},
Benedetta Gaia Varriale\IEEEauthorrefmark{2}
}

\IEEEauthorblockA{\IEEEauthorrefmark{1}
\textit{University of North Carolina at Charlotte}, Charlotte, NC, USA \\
d.cotroneo@charlotte.edu}

\IEEEauthorblockA{\IEEEauthorrefmark{2}
\textit{University of Naples Federico II}, Naples, Italy; \IEEEauthorrefmark{3}
\textit{IMT School for Advanced Studies Lucca}, Lucca, Italy \\
giuseppe.derosa20@unina.it,
cristina.improta@unina.it,
benedettagaia.varriale@unina.it}
}

\maketitle

\begin{abstract}

Understanding how software defects manifest and evolve in production environments is critical for improving reliability. While previous research has largely focused on pre-release defects, the nature of \emph{residual faults}, i.e., those escaping testing and surfacing post-release, remains poorly understood. This paper presents a large-scale characterization of pre- and post-release defects across C/C++ and Java systems, encompassing over 14k defects mined from open-source projects. We employ a broad suite of software metrics to capture diverse code attributes such as complexity, size, structure, and development history.

Results show that post-release defects are concentrated in older, frequently modified, and high-churn components, typically requiring longer and more complex fixes than pre-release ones. These findings highlight that residual defects arise more from evolutionary and process dynamics than code structure alone, suggesting that reliability engineering should prioritize targeted testing in mature and complex code regions.

\end{abstract}


\section{Introduction}
\label{sec:introduction}
Software failures in deployed systems remain a primary threat to dependability,
often arising from residual faults, i.e., bugs that evade rigorous testing
and manifest only in production~\cite{le1996ariane5, nasa1999mco}.
These latent defects have caused high-impact economic, operational, and
safety incidents~\cite{cnn2024crowdstrike, prokop2024softwareairplane}.

The urgency of understanding these escaping defects is increasing.
As the industry shifts toward AI-assisted development, analysts predict that
“by 2027, 25 percent of software defects escaping to production will result
from a lack of human oversight of AI-generated code”~\cite{gartner2024aiassistants}.
Yet, a critical gap persists: we still lack a fundamental understanding of
what properties allow certain bugs to escape testing, even in traditional,
human-authored software.

Current research largely treats software faults as a homogeneous category,
drawing on datasets dominated by easy-to-detect pre-release bugs~\cite{zhu2025bugs, thota2020survey}.
This introduces a survivorship bias: we design and optimize testing tools
based on the faults they already catch, rather than on those that evade them.
Without establishing a solid empirical baseline of how residual faults differ
from non-residual ones across product, process, and historical dimensions, 
efforts to secure future high-velocity development pipelines are proceeding blindly.

We address this gap by treating residual faults as a distinct class, isolating the specific characteristics that allow them to evade testing. This distinction is critical for reliability engineering: it shifts the focus from generic bug prevention to targeting the specific patterns, often invisible to standard metrics that manifest only in production. To this aim, we constructed a large-scale, balanced dataset from mature C/C++ and Java projects. Identifying residual faults is challenging due to their rarity and the noise in issue trackers; we filtered over $200k$ candidates to identify approximately $7k$ confirmed residual defects. We then reconstructed the faulty and fixed versions of each function, determining the escape timing by triangulating issue metadata, report text, and reporter identity.

To quantify differences between residual and non-residual faults, we extract a set of product, process, and statistical metrics of the code involved in each defect. Product metrics capture structural properties of the affected functions, while process metrics reflect their commit history, developer activity, and related aspects. We analyze these metric profiles to uncover systematic differences between pre-release and post-release faults, how these differences vary across languages, and relate to the complexity and effort of fixing each defect.

The contributions of this work are threefold.
\textit{(i)} We introduce a methodology for collecting, classifying, and reconstructing residual faults at the function level across heterogeneous systems.
\textit{(ii)} We construct two large-scale datasets enriched with product, process, and statistical metrics.
\textit{(iii)} We carry out a comprehensive evaluation and characterization of pre- and post-release defects. 

Overall, our analysis reveals three main findings:

\begin{enumerate}
\item Residual and non-residual defects are influenced by similar metric families, but the relationships among these metrics differ substantially: residual faults exhibit more heterogeneous and unstable metric profiles than defects detected during testing.

\item Distributional differences show that residual and non-residual defects diverge mainly along process and project-history dimensions, with structural metrics playing a smaller role, particularly in Java.

\item In C/C++, residual defects are not only harder to find but significantly more expensive to fix, requiring more complex code changes and time compared to pre-release bugs. In contrast, Java shows a more uniform fixing effort.

\end{enumerate}

To foster academic and industry research, we publicly release the datasets, the results, and code used in this work~\cite{RP}.

The remainder of this paper is organized as follows:
\S{}~\ref{sec:related} introduces background concepts and discusses previous work;
\S{}~\ref{sec:dataset} describes the dataset construction process and the software metrics used;
\S{}~\ref{sec:evaluation} reports the experimental evaluation;
\S{}~\ref{sec:threats} analyzes threats to validity;
\S{}~\ref{sec:conclusion} concludes the paper.

\section{Background \& Related Work}
\label{sec:related}

Software dependability research often uses the terms \emph{fault}, \emph{defect}, and \emph{bug} interchangeably. Following Laprie’s terminology~\cite{laprie1995dependability}, we refer to a \emph{fault} as a latent imperfection in the code or design that may lead to failure once activated, while we use \emph{defect} and \emph{bug} as synonyms commonly adopted in empirical software engineering. A \emph{failure} denotes the externally observable incorrect behavior that results from fault activation.

A \emph{release} represents a version of the software delivered to end users. Defects discovered before this point, in the testing phase, are \emph{pre-release defects} or \emph{non-residual}, whereas those that escape verification activities and manifest during operational use are \emph{post-release}, or \emph{residual} defects~\cite{laprie1995dependability, rwemalika2019industrial}. Unlike pre-release defects, residual defects tend to require specific runtime conditions to emerge~\cite{gazzola2017exploratory, dvorak2009nasa} and often result in reliability and economic consequences~\cite{rwemalika2019industrial, cnn2024crowdstrike}.

Research on software defects spans several directions, but only a limited portion directly addresses the post-release faults.

A first line of work investigates how software components accumulate pre- and post-release defects. Early studies show that defect patterns observed before release do not necessarily translate to failures in the field~\cite{fenton2002quantitative}. Other works highlight the role of code churn, historical change patterns, and product metrics in correlating with post-release defects across large industrial systems~\cite{nagappan2005explain, nagappan2006mining, nagappan2011code, zimmermann2007predicting}. These studies collectively indicate that post-release defects are influenced by both structural and historical properties but do not provide a direct characterization of how these faults differ from pre-release ones.

A broader body of research focuses on predicting post-release defect-proneness using product and process metrics. Systematic analyses show that process and change-related metrics often outperform traditional complexity-based ones for identifying defect-prone components~\cite{radjenovic2013software}. Later studies explored uncommon metric families such as cohesion measures and code-review characteristics~\cite{yang2014slice, krutauz2020code}. These approaches provide valuable predictors but aim at post-release prediction and rely on metric families that are costly to compute at scale.

Another research direction examines the conditions activating residual faults. Laprie~\cite{laprie1995dependability} emphasizes that certain faults activate only under specific environmental or workload factors not exercised during testing. Empirical analyses of field failures corroborate this view, showing that many faults leading to operational failures are triggered by rare execution scenarios or stress conditions, often require larger changes, and involve complex configuration interactions ~\cite{gazzola2017exploratory, sullivan1991software, rwemalika2019industrial}. These studies reveal behavioral properties of residual faults but do not compare them against pre-release defects.

Complementary work investigates the characteristics and representativeness of post-release faults. Studies using synthetic and real-world fault datasets highlight recurring semantic patterns, strong dependencies on execution context, and the difficulty of activating these faults during testing~\cite{duraes2006emulation, natella2013fault, cotroneo2025pyresbugs}. In Python, the majority of these faults are algorithm or interface\cite{cotroneo2025pyresbugs}, according to the ODC (\textbf{O}rthogonal \textbf{D}efect \textbf{C}lassification) scale ~\cite{chillarege1992orthogonal}. However, these studies are typically limited to specific ecosystems or languages and do not generalize across heterogeneous systems.

Overall, existing studies indicate that residual faults are structurally and behaviorally distinct, but systematic analyses quantifying these differences at scale are still rare. There is also no consensus on which metrics best capture these differences, with previous work highlighting both process and size-related dimensions. 
Our work addresses this gap by providing a qualitative characterization of residual defects using fine-grained code metrics and contrasting them with pre-release defects at scale.
Our findings offer a clearer understanding of how these faults arise in real systems, providing actionable insights for defect prediction, testing prioritization, and the design of more robust and representative techniques.


\section{Methodology}
\label{sec:dataset}
\begin{figure*}[ht!]
    \centering
    \includegraphics[width=0.75\linewidth]{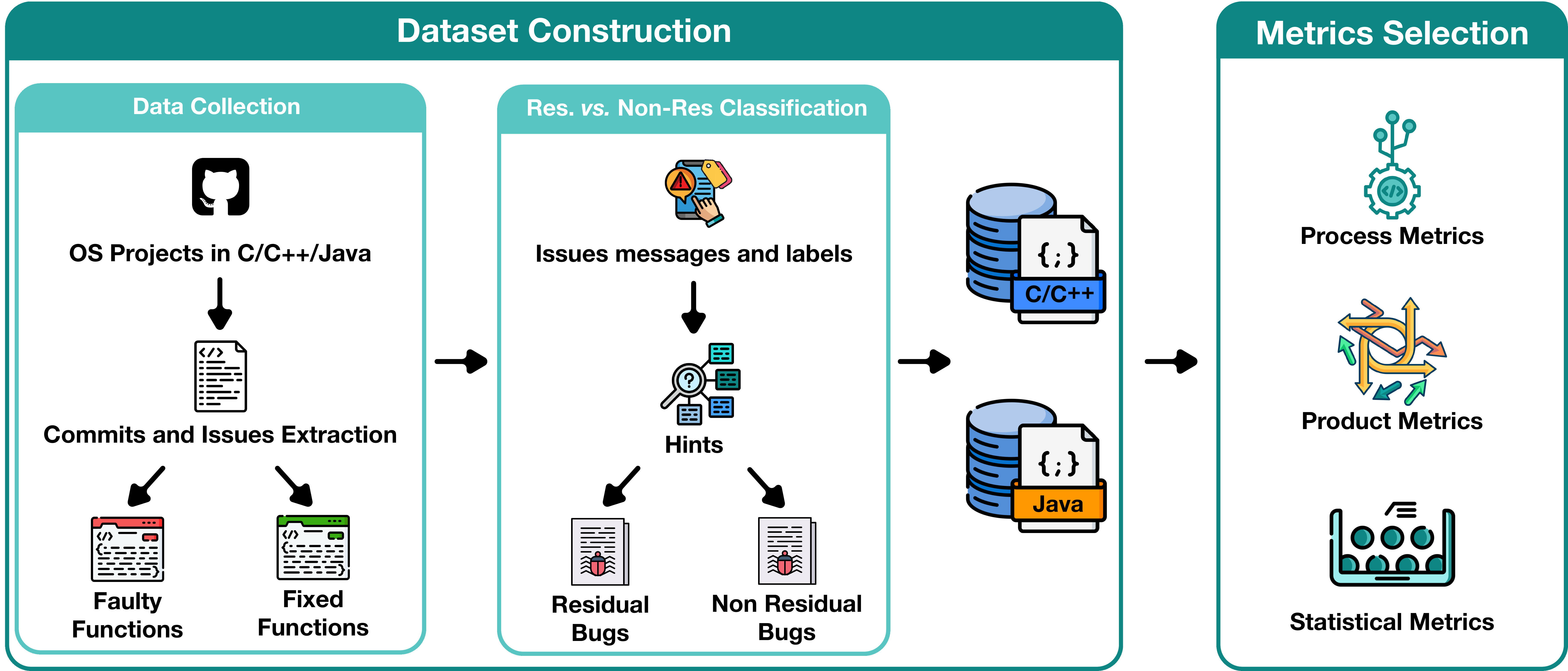}
    \caption{Overview of the methodology, including data collection and defects characterization.}
    \label{fig:methodology}
\end{figure*}

To characterize residual and non-residual defects, we construct a large-scale, representative dataset of defects drawn from real-world C/C++ and Java projects hosted on \textit{GitHub}, and for each faulty function and its corresponding fix, we extract a rich set of product, process, and statistical metrics that form the basis of our analyses.
\figurename~\ref{fig:methodology} shows an overview of the proposed methodology, illustrating the dataset construction process, including both data collection (\S{}~\ref{subsec:data_collection}) and defects classification (\S{}~\ref{subsec:heuristic}), and the large set of software metrics employed for our study (\S{}~\ref{subsec:metric_selection}).


\subsection{Data Collection}
\label{subsec:data_collection}

Characterizing residual defects in comparison with non-residual ones requires more than a generic dataset of bugs. For each defect instance we need \textit{(i)} a reliable distinction between non-residual (i.e., pre-release) and residual (i.e., post-release) defects, \textit{(ii)} an explicit mapping between the faulty and fixed versions of the code at function level, and \textit{(iii)} enough projects and diversity to make the results meaningful across languages and representative of real-world systems. Existing datasets of real defects only partially satisfy these requirements. Publicly available datasets such as Defects4J~\cite{rene2014defects4j}, ManyBugs~\cite{legoues2015manybugs}, PyResBugs~\cite{cotroneo2025pyresbugs} and other popular ones~\cite{saha2018bugsjar, tomassi2019bugswarm, jimenez2024swebenchlanguagemodelsresolve}, provide valuable collections of real-world defects. Still, they differ in granularity, language coverage, and the type of artifacts they expose. In particular, they typically do not offer precise project-level metadata, provide inconsistent or coarse mappings between faulty and fixed code fragments, and do not systematically separate pre-release from post-release defects across a broad and diverse set of mature projects.

To overcome these limitations, we construct two new datasets by mining real-world GitHub repositories, focusing on C/C\texttt{++} and Java as target languages, since they are both widely used in reliability-critical systems, represent two different programming paradigms (e.g., managed \textit{vs.} unmanaged memory), and offer large, mature open-source ecosystems with rich development histories and issue-tracking data. Our goal is to obtain a large-scale, function-level view of both residual and non-residual defects from mature projects, while preserving all the information needed for their characterization.

For each defect instance, we record \textit{(i)} \textit{repository and commit metadata}, including the URL of the GitHub repository, the SHA (i.e, Secure Hash Code, a unique identifier of each commit) of the fixed commit, and the path of the affected file in the repository, 
\textit{(ii)} \textit{change level information}, namely the diff patch 
capturing all modifications between the faulty and fixed versions, and \textit{(iii)} \textit{code level information}, that is, the faulty and fixed code snippets at function level together with the corresponding function or method name. 
These fields provide all the context needed to compute process, product, and statistical metrics for the characterization of defects (as detailed in \S{}~\ref{subsec:metric_selection}).

To populate these fields, we start from a curated list of  C/C++ and Java repositories and analyze the complete commit 
history of each project to identify bug-fixing commits. Following common practice in mining software repositories, we rely on a keyword-based heuristic on commit messages~\cite{sliwerski2005whendochanges}. A commit is considered a bug-fixing commit if its message contains at least one of the keywords \texttt{fix}, \texttt{bug}, \texttt{resolve}, or \texttt{patch}. This simple heuristic is widely used and achieves strong precision for large-scale studies~\cite{campos2017mining, yu2019characterizing}.

For each commit classified as bug-fixing, we mark it as the \emph{fixed commit}. This corresponds to the assumption that the bug-fixing commit introduces the repaired code, while the parent commit still contains the faulty version. We then invoke \texttt{git diff} on the fixed commits and store the resulting patch. The patch contains the paths of modified files and the line-level edits later used to locate the faulty and fixed code snippets.

Identifying which specific function or method is affected by each bug is crucial for our characterization and for excluding trivial edits that do not represent meaningful defects (e.g, adding a logging or print statement). To this end, we developed a dedicated pipeline that operates directly on the repository history and parseable code, so that we can focus on substantive behavioral changes in the affected functions.

First, for each diff hunk in the patch of a bug-fixing commit, we scan the diff context to identify function or method declarations surrounding the changed lines, yielding one or more candidate function names likely impacted by the fix.

Second, we clone the repository and check out both versions of the affected file: the faulty version (parent of the bug-fixing commit) and the fixed version (the bug-fixing commit), as identified by their SHAs. We parse both versions with \texttt{tree-sitter}~\cite{treesitter} for C/C++ and Java to obtain abstract syntax trees, from which we extract all function or method definitions with their names and bodies. We then match the candidate names from the diff against these definitions and, when a match is found, extract the entire function body from the faulty file as the faulty snippet, and the corresponding body from the fixed file as the fixed snippet. This gives us function-level faulty–fixed pairs suitable for computing product, process, and statistical metrics~\cite{tufano2019codechanges,tufano2019learning}. We additionally perform a syntactic check on all extracted functions to ensure that each snippet is complete and well-formed.

After extracting faulty and fixed snippets for all bug-fixing commits, we link each commit to its corresponding issue (i.e, an issue is a discussion item in the project issue tracker used to report bugs, feature requests or discuss development tasks) in the project tracker whenever such a link exists, using standard patterns such as issue identifiers mentioned in commit messages and explicit GitHub links. This issue-level information, combined with repository release tags and commit dates, serves as input to the heuristic that classifies each defect as residual (i.e., defects that escaped testing and were fixed only after release) or non-residual (i.e., defects caught during testing and fixed before release).




\subsection{Residual \textit{vs.} Non-residual Defects Classification}
\label{subsec:heuristic}

Issue trackers are noisy, and issue types are often misclassified, so relying on a single field, such as the issue label, is not sufficient to distinguish residual from non-residual defects. We therefore design a novel multi-signal heuristic that combines information from the issue labels, issue body, and reporter identity. Each source contributes hints on whether a given issue is a defect or not and, for defects, whether it is more likely to correspond to a post-release or a pre-release defect. We apply the heuristic to all issues linked to bug-fixing commits and then aggregate the signals to obtain a final classification.

We implement a fully automated, reproducible heuristic classifier based solely on local signals.
\circled{1} First, we inspect the \textit{labels} (i.e, a label is a user-defined tag attached to an issue that helps categorize its content) associated with each issue. Labels such as \textit{bug}, \textit{defect}, \textit{bugfix}, and \textit{regression} are treated as strong indicators that the issue is a defect. While labels such as \textit{enhancement}, \textit{feature request}, \textit{proposal} or \textit{idea} are interpreted as signals that the issue is a feature or improvement request rather than a defect. Similarly, labels that explicitly refer to \textit{question}, \textit{support}, \textit{usage}, \textit{documentation} or \textit{discussion} are interpreted as signals that the issue does not correspond to a defect at all. Using issue labels in this way follows common practice in issue classification and defect prediction studies~\cite{abedini2024githubissues, herbold2020feasibility}. At this stage, we distinguish defects from non-defects, but labels alone do not yet determine whether a defect is residual or not.

\circled{2}
For issues that remain ambiguous after label analysis, we perform a structured classification of the \textit{issue body} to determine whether the issue corresponds to a residual (\textit{post-release}), a non-residual (\textit{pre-release}) defect, or a non-defect.

First, we search for textual patterns that are characteristic of user-submitted bug reports. These include sections or expressions such as \textit{issue description}, \textit{steps to reproduce}, \textit{expected behavior}, \textit{actual behavior}, \textit{environment}, \textit{configuration}, \textit{operating system}, and \textit{Java version}. Such elements are commonly embedded in issue templates and typically indicate that the defect was observed during operational use. When these structured bug-report signals are present, we classify the issue as a residual defect.

Second, we identify signals that the issue was discovered during internal validation activities. References to continuous integration artifacts (e.g., \textit{CI link}, \textit{continuous integration}, \textit{test failure}), testing frameworks (e.g., \textit{JUnit}, \textit{assertj}, \textit{assertEquals}, \textit{assertThat}, \textit{AssertionError}), or development branches (e.g., \textit{dev}, \textit{develop}, \textit{main}, \textit{master}, \textit{trunk}) indicate that the defect was detected during pre-release testing. In such cases, we classify the defect as non-residual.

Third, we detect expressions that suggest enhancement proposals or maintenance activities rather than defect reports, such as \textit{it would be great if}, \textit{proposal for}, \textit{feature request}, \textit{cleanup}, \textit{refactor}, \textit{work in progress}, and \textit{TODO}. Issues dominated by this language are classified as non-defects.

Finally, issues primarily centered on usage questions, configuration misunderstandings, documentation clarification, or general support topics are also classified as non-defects and excluded from further analysis.

\begin{table}[t]
\centering
\caption{Dataset Statistics}
\begin{tabular}{lcccccc}
\toprule
\textbf{Language} & 
\multicolumn{2}{c}{\textbf{Samples}} & \textbf{Total Non-Residual}\\
 & Residual & Non-Residual \\
\midrule
C/C++  & 4010 & 4010 & 96541 \\
Java   & 3118 & 3118 & 89851  \\
\midrule
\textbf{Total} & 7128 & 7128 & 186,392 \\
\bottomrule
\end{tabular}
\label{tab:dataset}
\end{table}

\circled{3} If labels and issue text do not provide sufficient evidence, we consider the \textit{type of reporter}. Issues reported by external contributors, such as users who are not members or maintainers of the project, are more likely to correspond to residual, post-release defects, since external users typically interact with released versions of the software. Conversely, issues opened by project members or core maintainers are more likely to correspond to internal defects discovered during development or testing, and are therefore treated as weak signals of pre-release defects. Reporter identity is used only as a secondary, tie-breaking signal, and added to the other, since external contributors can also report pre-release issues in some settings.

\circled{4} For each issue, we aggregate all signals from labels, issue body, and reporter type. If all activated signals support the same category, the issue is assigned to that category. If multiple conflicting signals appear, or no rule is triggered, the issue is conservatively labeled as \textit{unknown} and excluded from the dataset.
This conservative strategy avoids forced decisions in ambiguous cases, reduces the risk of mislabeling, and prioritizes precision over recall in the identification of residual defects.


The final collected dataset comprises $193k$ samples overall, including $4k$ and $3k$ residual defects, and $96k$ and $90k$ non-residual defects for C/C++ and Java, respectively. 


Subsequent analyses are conducted on a balanced subset of the collected data to ensure that residual and non-residual defects are randomly selected from the same software projects. This design choice helps control for domain complexity and ensures that observed differences are not driven by unrelated project-level variation. By focusing on defects within the same codebases, we isolate factors that are inherent to the nature of the faults themselves and to the characteristics of the code in which they occur.
Detailed statistics are presented in \tablename~\ref{tab:dataset}, including the \textit{total} number of non-residual faults present in the dataset and the balanced subset used for the analyses. 

To ensure the reliability of our classification procedure, we complemented the automated heuristic with a manual validation. A group of two Ph.D. students in computer science, together with a postdoctoral researcher holding a Ph.D. in information technologies, examined a statistically representative sample of 400 issues, sized to achieve a 5\% margin of error. Each item was independently checked to verify whether the heuristic’s decision matched the interpretation derived from a full reading of the labels, textual description, and contextual project information. Out of the 400 inspected cases, only 12 were found to be misclassified (3.0\%), confirming that the heuristic achieves a high level of agreement with expert judgment.
Disagreements were discussed until a consensus was reached. In case of no consensus, the subset of uncertain cases was further reviewed by a senior faculty member with extensive expertise in software engineering. The combined effort ensured that the heuristic behaves as intended and that its decisions remain aligned with expert judgment, strengthening the validity of the resulting dataset.

All stages of the data collection and classification pipeline are fully automated. Repository mining, commit extraction, diff generation, function-level reconstruction, issue linking, and multi-signal classification are built on top of git commands and the GitHub REST API. 
Although our implementation targets GitHub repositories, the methodology is designed around platform-agnostic concepts common to modern software repositories, including version control history, commit metadata, issue tracking records, release tagging, and textual issue descriptions. The classification logic operates on generic signals such as defect labels, workflow references, reporter roles, and release timing, which are available in platforms such as GitLab, Bitbucket, and other issue-tracking systems. 
We publicly share the code and complete dataset for transparency and replication purposes~\cite{RP}.

\subsection{Software Metrics Extraction}
\label{subsec:metric_selection}

\begin{table*}[htbp]
\centering
\scriptsize
\caption{Software Metrics adopted in our study.}
\begin{tabularx}{\textwidth}{>{\raggedright\arraybackslash}p{1.6cm} >{\arraybackslash}p{2.3cm} X}
\toprule
\textbf{Category} & \textbf{Sub-Category} & \textbf{Metrics} \\
\midrule

\multirow{9}{*}{\textbf{Product} (22)}
  & \multirow{2}{*}{\centering Complexity} 
    & Halstead Difficulty, Halstead Vocabulary, Halstead Volume, Halstead Effort,
      Halstead Total Operators, Halstead Total Operands, Halstead Distinct Operators, Halstead Distinct Operands, Cyclomatic Complexity, Count of Paths \\
\addlinespace

  & Coupling 
    & Component Reuse, Fan-In, Fan-Out \\
\addlinespace

  & Documentation 
    & Blank Lines of Code, Comment Lines of Code, Ratio Comment to Code \\
\addlinespace

  & Inheritance 
    & Depth of Inheritance Tree \\
\addlinespace

  & \multirow{1}{*}{\centering Size} 
    & LOC, Count of Statements, Count of Declaration Statements,
      Count of LOC Declarations, Count Lines of Executable Code \\
\addlinespace

\midrule

\multirow{8}{*}{\centering\textbf{Process} (21)}
  & Temporal 
    & Age in Days, Avg. Code Churn, Max. Code Churn, Total Code Churn \\
\addlinespace

  & \multirow{2}{*}{\centering Commit-Based}
    & Commit Churn Added, Commits Count, Max. Commit LOC Changes,
      Avg. Commit LOC Changes, Total Commit Churn, \\
  & & Commit Churn Deleted, Change Pattern Count, Total Commits, Bug Density \\
\addlinespace

  & \multirow{1}{*}{\centering Method-Based}
    & Total Method Churn, Method Churn Added, Method Churn Deleted,
      Avg. Method LOC Changes, Max. Method LOC Changes \\
\addlinespace

  & Developer-Based
    & Number of Distinct Authors, Avg. Developer Experience,
      Avg. Commits per Author \\
\addlinespace

\midrule

\textbf{Statistical} &  & Entropy \\
\bottomrule
\end{tabularx}
\label{tab:metrics}
\end{table*}


The goal of this study is to characterize residual defects and understand how they differ from non-residual ones. To the best of our knowledge, prior work does not identify a clear set of distinguishing properties for these faults. To capture the most relevant information about them, we conducted a literature review on software metrics that are effective in modeling or predicting software faults~\cite{radjenovic2013software, li2018progress,li2024software}.

From this review, we derived a pool of 44 candidate metrics, which we organized into three main groups: \emph{product metrics} describe static characteristics of the source code; \emph{process metrics} capture historical and evolutionary aspects of the codebase; and \emph{statistical metrics}, such as entropy, quantify how natural or irregular the distribution of code tokens appears. 
Together, these metrics cover complexity, size, coupling, documentation, change history, developer activity, and statistical properties of the code that may support the discovery of residual defects.
\tablename~\ref{tab:metrics} reports the complete list of metrics considered in our study.

\textbf{Product Metrics.} These metrics capture structural properties of the code that are known to influence defect proneness and testability. In our study, they include metrics that assess \textit{complexity} (e.g., Halstead difficulty, vocabulary, volume, effort, cyclomatic complexity), \textit{coupling} (e.g, component reuse, fan-in), \textit{documentation} (e.g., comment-to-code ratio), \textit{inheritance depth}, and \textit{size} (e.g., lines of code, number of statements). Together, these quantify how large, intricate, and interconnected a function is, as well as how well it is documented. We use them to investigate whether residual faults tend to occur in structurally more complex, larger, more tightly coupled, or more poorly documented code than non-residual faults, and to what extent such structural differences are consistent across languages.

\textbf{Process Metrics.}
These metrics complement the structural perspective by characterizing how the code has evolved over time.
\textit{Temporal} metrics (e.g., age, code churn) describe the lifetime and volatility of a function.
\textit{Commit-based} metrics capture how changes are distributed across the commits that modify the file containing the fix (e.g., commit churn added, number of commits).
\textit{Method-based} metrics instead quantify the evolution of the specific method under analysis by aggregating the modifications performed across all commits that touched it (e.g., maximum method LOC changes).
Finally, \textit{developer-based} metrics summarize properties of the contributors who modified the code (e.g., distinct authors, avg. experience). These metrics allow us to investigate whether residual defects are more likely to appear in older, heavily modified projects, and in code touched by many or relatively inexperienced developers, compared to non-residual faults. 

\textbf{Statistical metrics.}
In addition to product and process metrics, we consider a statistical, entropy–based metric that captures how ''natural'' a code snippet is. Code naturalness refers to the tendency of source code to exhibit repetitive, predictable patterns that can be learned by statistical language models: a snippet is considered more natural if a model trained on code assigns it a higher probability (hence, lower entropy) \cite{hindle2016naturalness,rahman2019natural}. Naturalness has also been tied to software quality: faulty lines tend to be less natural than non-faulty or fixed lines and are therefore identifiable by language models \cite{ray2016naturalness}. 

We leverage this notion to characterize residual and non-residual defects from a probabilistic perspective, assessing whether residual defects tend to occur in code that is systematically more or less natural than the code hosting non-residual defects.
Concretely, we compute for each code sample a cross-entropy score under an $n$-gram language model of source code. Higher entropy in this setting indicates code that is less expected under the model and thus less natural, providing a quantitative proxy for the statistical irregularity of the locations where defects appear.


Following best practices in the field~\cite{hindle2016naturalness, rahman2019natural}, we remove comments, tokenize code and apply role-aware identifier normalization. Then, the corpus is randomly partitioned into $k$ folds. For each fold, we train an $n$-gram language model using only the normalized fixed-code functions in the training partition for that fold. Training on fixed (i.e., the faulty-fault-free version of the code) code allows the model to capture the regularities, idioms, and stylistic conventions of correct code, as done in previous studies that compare buggy and fixed code using entropy \cite{jiang2025empirical, ray2016naturalness}. Using cross-validation 
yields unbiased estimates of how natural each code sample is.

Once a model has been trained for a given fold, we use it to score all faulty samples in the corresponding test partition. 
The model provides a log-probability for each token, which we then aggregate per line and convert into a cross-entropy score (in bits) using the standard formulation~\cite{bishop2006}. 


Each sample consists of multiple lines and therefore multiple entropy values. Rather than averaging over all lines, which would dilute the contribution of highly surprising lines, we summarize the distribution of line entropies for a sample by its 90th percentile. This summary focuses the metric on the most entropic portion of the snippet while remaining robust to single-line outliers. 


\section{Evaluation}
\label{sec:evaluation}
Our evaluation is guided by the following research questions, which aim to investigate how residual defects differ from non-residual defects in ways that are measurable on source code and development history, and that can ultimately inform testing, defect prediction, and reliability engineering, as well as to understand whether these defect characterizations are consistent across programming languages.

\noindent
\textbf{RQ$_1$:} \emph{Which software metrics are most influential in characterizing pre-release vs. post-release defects across languages?}

\noindent
This RQ focuses on the joint profiles of product, process, and statistical metrics associated with faulty functions. To answer this, we first cluster metrics to see which ones move together and how their correlation structure differs between residual and non-residual faults, then apply PCA (\textbf{P}rincipal \textbf{C}omponent \textbf{A}nalysis) to  identify the main structural and process dimensions that explain most variance in each subset. 
Identifying influent and recurring metric profiles that are strongly associated with residual defects can guide the design of targeted testing and review strategies that concentrate effort on functions whose profiles resemble residual defects.

\vspace{0.1cm}
\noindent
\textbf{RQ$_2$:} \emph{Do software metrics profiles statistically differ between pre-release and post-release defects across languages?}

\noindent
We statistically compare the distributions of metrics 
between pre- and post-release defects using non-parametric tests and effect sizes to quantify the magnitude and practical relevance of the observed differences. 
If certain metrics show strong and consistent distributional shifts for residual defects across languages, they can serve as robust indicators for risk assessment, lightweight screening during code review, or feature selection in defect prediction models.

\vspace{0.1cm}
\noindent
\textbf{RQ$_3$:} \emph{What are the differences in effort and complexity when fixing pre-release vs. post-release defects across languages?}

\noindent
Residual defects are often harder to understand and repair once discovered. In RQ3, we compare the faulty and fixed code versions, along with their associated changes, to quantify structural variations (e.g., size and complexity) and effort-related aspects (e.g., scope and timing of fixes). This analysis informs both reliability engineering and project management. If residual defects consistently require larger, more complex, and longer fixes, this provides quantitative evidence of their higher maintenance cost. Such findings support greater investment in preventive measures, improve cost and release planning, and help prioritize post-release issues likely to demand significant effort and coordination.

\smallskip

To compute product metrics, we relied on \textit{Understand}~\cite{understand}, a certified, industry-level and widely employed code analysis tool. It supports a wide range of programming languages enabling us to compute product metrics in both datasets in a consistent and comparable way.  
Understand creates a database by parsing all source files. It constructs a detailed code model that captures the project's structure, symbol definitions, control flow and dependencies. 
As for process metrics, we used git commands to reconstruct the full evolutionary history of each method, extracting commit metadata, authorship information, and line-level modifications \cite{hassan2008road}. 
Finally, to compute the entropy-based metric, we trained \textit{KenLM}~\cite{heafield2011kenlm} as a $n$-gram language model, a scalable toolkit that has become the de facto standard for scalable entropy-based analyses. KenLM provides highly efficient estimation and querying also on large corpora, making it well-suited for software datasets with millions of tokens. 
Based on empirical testing, we found that the $n$-gram model stabilized in capturing local code patterns around $n{=}5$, as also supported by previous work~\cite{rahman2019natural}. 
We used 5-fold cross-validation, a standard choice in code modeling literature. 

All code for metrics computation is available in our replication package for transparency and reproducibility~\cite{RP}.

\subsection{Which software metrics are most influential in characterizing pre-release vs. post-release defects across languages?}

To investigate how software metrics relate to the nature of defects, we applied Ward's hierarchical clustering on residual and non-residual defects, separately, for both C/C++ and Java datasets. Our goal was to identify how correlations among metrics shift between functions containing pre-release and post-release defects, potentially revealing systematic behavioral distinctions.
First, we scaled data to reduce the impact of outliers, a fundamental step given the heterogeneity and skewed distributions commonly found in software metrics. Then, we computed pairwise distances between metrics based on their correlations. The resulting clusters help identify groups of metrics that behave similarly and uncover how those relationships change depending on the type of defect.

\begin{figure*}

    \centering
    \includegraphics[width=0.75\linewidth]{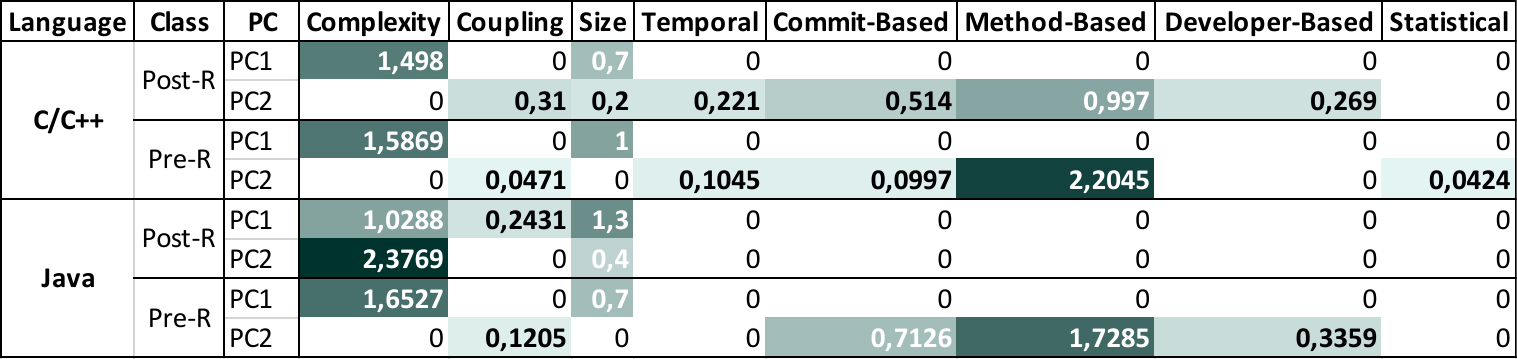}
    \caption{Comparison of principal components loadings per metric category.}
    \label{fig:pca_table}
\end{figure*}

In the C/C++ dataset, hierarchical clustering indicates that the relationships among metrics change noticeably between non-residual and residual defects, even though the underlying metrics are the same. For non-residual defects, executable lines and Halstead volume and effort measures form a single, coherent family, which suggests a unified ``size/effort'' dimension where syntactic size and Halstead-based complexity co-vary tightly. In the residual subset, this structure fragments: executable lines cluster with statement counts in a syntactic-size group, the Halstead volume metrics form a separate cluster, and Halstead effort detaches from both, indicating that syntactic size and volume no longer align as closely for faults that escape testing. 
Process metrics show a similar reorganization. Method-level churn measures that form a single cluster for non-residual defects split into different groups for residual defects, with maximum change joining added and total churn, and average change and deletions becoming more independent. 
At the same time, statement and executable-line counts, which are separated in non-residual defects, cluster together for residual ones. This reinforces the picture of a different internal correlation structure among size, complexity, and change metrics depending on whether a defect is caught before or after release.

Java defects exhibit a similar reorganization. For non-residual defects, \textit{LOC} and Halstead volume metrics form a single size/volume family, while vocabulary-related Halstead measures and effort behave as partially distinct dimensions; in this case, simple size and volume are almost interchangeable, vocabulary richness is related but separate, and effort is relatively independent. In the residual subset, all Halstead metrics, including vocabulary and effort, collapse into a single cluster, whereas line-based size forms its own group, indicating that, in functions hosting post-release defects, Halstead’s components jointly describe a unified notion of complexity that is no longer captured by line counts alone.
Process metrics also shift: avg. and max. method-level change magnitudes, which appear in separate clusters for non-residual defects, cluster together in the residual subset, indicating that edits become more tightly coupled in code associated with residual faults. 

\emph{Overall, in both languages, clustering shows that the way structural, churn, and commit-related metrics relate to one another systematically changes for defects missed by pre-release quality assurance. This supports the hypothesis that residual defects tend to appear in code regions with more complex, less predictable, or decoupled metric profiles, characteristics that may help explain why these faults are more difficult to detect.}

After the clustering analysis, we further examined which metrics account for the largest share of variability in each subset by applying Principal Component Analysis (PCA)~\cite{abdi2010principal} separately to non-residual and residual defects. PCA is a dimensionality reduction technique that transforms the original, potentially correlated metrics into a new set of orthogonal (``\textit{principal}'') components, each defined as a linear combination of the original variables and ordered by the amount of variance it explains. By inspecting the \textit{loadings} of the original metrics on the principal components, we identify the dominant structural and process dimensions in the data. 

\figurename~\ref{fig:pca_table} shows a heatmap that aggregates PCA loadings by metric category, with rows for each language (C/C++ and Java), defect type (pre-release and post-release), and principal component (\textit{PC1}, \textit{PC2}), and columns for the different metric categories (complexity, coupling, size, temporal, commit-based, method-based, developer-based, statistical). Categories of metrics that are not present in the first two components are omitted.
Here, we focus on the first two principal components (\textit{PC1} and \textit{PC2}) for brevity, which together explain $\sim$50-60\% of the total variance. 
The replication package contains complete results~\cite{RP}.

For non-residual C/C++ defects, \textit{PC1} captures a clear structural dimension dominated by size and Halstead volume metrics. All major Halstead measures (length, operators, volume, vocabulary) and line-based metrics (\textit{LOC}, executable lines) load strongly and move together, forming a unified ``size/volume'' axis. \textit{PC2} reflects process activity, driven mainly by method-level churn metrics (total, added, deleted, and maximum churn), while metrics like age, fan-out and entropy contribute marginally. It therefore represents the intensity of code modification rather than structural complexity. 

Residual C/++ defects show a broadly similar first component, again defined by size and Halstead metrics, but with statement-level measures (i.e., counts of statements and of executable code) contributing more strongly. The second component, however, diverges substantially. For residual defects, \textit{PC2} integrates coupling, change intensity, commit history, and commenting activity, with large loadings for fan-out, number of authors, total commits, age, bug density, and comment lines. In contrast to the nearly single-axis churn component observed for non-residual defects, the residual subset exhibits a more heterogeneous process dimension linking change behavior, coupling, and maintenance history.

The Java results follow a closely related pattern. For non-residual Java defects, \textit{PC1} (44.5\% variance) reflects a strong structural dimension dominated by Halstead and size metrics. Measures of length, operands, operators, volume, vocabulary, and line-based size (\textit{LOC}, executable lines, declaration statements) all load heavily, confirming that structural size and Halstead volume define a unified factor, as in C. \textit{PC2} (15.3\%) represents process activity, with high loadings for total, added, and deleted method churn, total commits, and max. method change, while distinct authors and fan-in contribute modestly. Thus, non-residual Java defects show a similar structural–process separation, where the second component captures the intensity of method-level churn.

For residual Java defects, \textit{PC1} remains structural but integrates statement-level measures more tightly. Halstead size and volume, line-based size, and statement counts load together, forming an even stronger unified ``code volume'' dimension. \textit{PC2} differs more substantially: it combines process, coupling, and historical factors, with large loadings for fan-out, number of authors, commits, age, churn measures, and commenting. As in residual C, this indicates a broader and more heterogeneous process dimension where coupling, change history, and fault concentration jointly characterize residual defects, rather than method churn alone.

Comparing across languages, three observations emerge. First, \textit{PC1} is always a structural factor dominated by size and Halstead volume metrics. Regardless of defect type and language, the largest share of variance in faulty functions is explained by structural size and volume. Second, in both languages \textit{PC2} for non-residual defects is primarily a method-churn and commit-history factor, whereas for residual defects it becomes a richer process–architecture factor that mixes churn, commit history, coupling, age, fault density, and commenting. This shift is consistent across C/C++ and Java and indicates that the process and architectural context of residual defects is more complex than that of defects caught pre-release. Third, statement-level metrics (e.g., counts of statements and declarations) contribute more strongly to the leading structural component in the residual subsets than in the non-residual subsets, suggesting that the syntactic organization of code at the statement level is more tightly bound to size and volume in locations that host residual defects.


\begin{finding}{}{}
\textbf{Key Finding 1:} \emph{In both C/C++ and Java pre-release and post-release defects are characterized by the same core structural and process metric families (size/volume, complexity, churn, history, coupling). However, residual faults are characterized by a systematically different correlation and variance structure in which size and Halstead dimensions decouple from simple line counts and the main process-level metrics component shifts from ``pure churn'' to a richer combination of churn, commit history, coupling, age, and fault concentration.}
\end{finding}{}

\subsection{Do software metrics profiles statistically differ between pre-release and post-release defects across languages?}

Understanding whether software metrics exhibit systematically different distributions for residual and non-residual defects is essential to determine which metrics meaningfully characterize faults that escape pre-release testing. Distributional differences would indicate that post-release defects tend to arise in code regions with distinct profiles.

We followed a four-step analysis pipeline: \textit{(1)} inspect the empirical distributions of all metrics; \textit{(2)} apply a two-sample statistical test; \textit{(3)} quantify the magnitude of the observed differences; and \textit{(4)} compare trends across languages. Because the metric distributions computed on both C/C\texttt{++} and Java datasets are usually non-normal, skewed, and heavy-tailed, we employ the \textit{Kolmogorov--Smirnov} (KS) test~\cite{massey1951kolmogorov}, a non-parametric test that compares two distributions without assuming any specific form. The null hypothesis ($H_0$) states that residual and non-residual defects are drawn from the same distribution; the alternative ($H_1$) states that they differ. The identified difference exists anywhere in the distribution, regardless of whether it arises in the location, scale, skewness, or tails. 
To quantify the practical magnitude of these differences, we report four complementary indicators. The KS statistic measures the largest distance between the empirical cumulative distributions of residual and non-residual defects, capturing how strongly the two distributions diverge. The associated $p$-value tests the null hypothesis that the two samples come from the same distribution. We also compute Cliff’s $\delta$~\cite{grissom2005effect}, which estimates the probability that a randomly chosen residual defect has a higher (or lower) metric value than a non-residual one. Finally, we classify the Magnitude of each effect (negligible, small, medium, large) to contextualize the strength of the observed difference. Table~\ref{tab:ks-effects} reports all metrics exhibiting at least a \textit{small} effect size in either dataset.

In C/C++, 35 out of 36 (97\%) metrics exhibit statistically significant differences between residual and non-residual defects ($p<0.05$), but only 11 reach at least a small effect size according to Cliff’s $\delta$. The strongest shift occurs for \textit{age days} (i.e., function existence in the project in days), which shows a \textit{large} effect ($\delta = 0.63$), indicating that residual defects arise in substantially older code. Several process and history metrics show \textit{medium} effects. A further group of metrics, like \textit{fan-in} or \textit{component reuse}, exhibits \textit{small} but consistent effects. The remaining metrics (e.g., \textit{LOC}, Halstead measures, nesting, comment ratios) often have extremely small KS $p$-values but negligible effect sizes, suggesting that while their distributions differ in a statistically detectable way, the magnitude of the shift is modest in practice.

In Java, 30 out of 36 (83\%) display statistically significant KS $p$-values, but only two cross the small-effect threshold: \textit{age days}, with a \textit{medium} effect ($\delta = 0.39$), and \textit{distinct authors}, with a \textit{small} negative effect. As in C/C++, residual defects tend to be associated with older components, but in Java, they are concentrated in files touched by fewer distinct authors. 

\begin{table}[t]
\centering
\caption{KS statistics and effect sizes (Cliff’s $\delta$) for metrics showing at least a \emph{small} effect when comparing residual vs.\ non-residual defects in C/C++ and Java.}
\label{tab:ks-effects}
\scriptsize
\begin{tabular}{lrrcr}
\toprule
\multicolumn{5}{c}{\textbf{C/C++ Dataset}}\\
\midrule
\textbf{Metric} & \textbf{KS} & \textbf{$p$-value} & \textbf{$\delta$} & \textbf{Magnitude} \\
\midrule
Age in days & 0.6170 & 0                          & $|0.6274|$ & large \\
Bug density & 0.2853 & 3.9e-137                & $|0.3571|$ & medium \\
Method churn added & 0.3088 & 6.0e-161        & $|0.3497|$ & medium \\
Total commits & 0.2911 & 7.8e-143              & $|0.3458|$ & medium \\
Total method churn & 0.3074 & 1.7e-159        & $|0.3493|$ & medium \\
Max. method LOC changes & 0.2835 & 2.0e-135   & $|0.3148|$ & small \\
Avg. method LOC changes & 0.2432 & 1.9e-99    & $|0.2641|$ & small \\
Distinct authors & 0.1741 & 6.6e-51            & $|0.1487|$ & small \\
Component Reuse & 0.1606 & 1.5e-45               & $|0.1647|$ & small \\
Fan-In & 0.1948 & 1.2e-58                      & $|0.2128|$ & small \\
\midrule
\multicolumn{5}{c}{\textbf{Java Dataset}}\\
\midrule
\textbf{Metric} & \textbf{KS} & \textbf{$p$-value} & \textbf{$\delta$} & \textbf{Magnitude} \\
\midrule
Age in days & 0.3031 & 1.6e-126 & $|0.3856|$ & medium \\
Distinct authors & 0.2462 & 4.0e-83 & $|0.3057|$ & small \\
\bottomrule
\end{tabular}
\end{table}

All other structural and process metrics show negligible effect sizes.
Specifically, in C/C++, the statement- and comment-level metrics (e.g., \textit{counts of statements and of lines of executable code}) show significant KS differences, whereas their Java counterparts do not. This suggests that the size and documentation metrics are more strongly associated with residual defects in C/C++ than in Java.

To further validate the statistical findings, we trained supervised models to assess the discriminative strength of the metrics in a multivariate setting. In both languages, XGBoost~\cite{chen2016xgboost} achieved substantially high predictive performance (C/C++ F1: 0.91; Java F1: 0.86), confirming the presence of meaningful distributional differences. Importantly, the top contributors identified by XGBoost, such as age, reuse and dependency measures, authorship patterns, churn indicators, and bug density, align closely with the metrics showing the largest KS effects. This convergence between univariate statistical tests and multivariate predictive models strengthens the robustness of our conclusions.

\begin{finding}{}{}
\textbf{Key Finding 2:} \emph{Several metrics differ statistically between pre-release and post-release defects, but only a subset exhibits non-negligible effect sizes. Across both languages, the most evident differences arise from process and history metrics: age, churn, bug density, reuse, authorship, and fan-in. Traditional structural metrics play a secondary role, especially for Java. Residual defects are thus more likely to occur in components that are older, more frequently modified, and embedded in richer change and dependency histories.}
\end{finding}{}

\subsection{What are the differences in effort and complexity when fixing pre-release vs. post-release defects across languages?}

Assessing the effort required to resolve pre- and post-release defects provides valuable insight into the practical consequences of residual faults. As discussed earlier, residual defects are typically harder to diagnose and escape traditional testing processes, often surfacing only under real-world execution conditions. Because these faults tend to incur higher reliability and maintenance costs, analyzing the effort and complexity associated with their resolution offers a concrete way to quantify their additional burden.

The goal of this research question is to determine whether residual (post-release) defects demand greater human and technical effort to fix than non-residual (pre-release) ones. To address this, we investigate both effort-based and complexity-based dimensions of defect fixing. The first captures the human and temporal cost of resolving defects, while the second quantifies the structural and cognitive changes introduced in the code during the fix.

\begin{figure}[ht]
    \centering

    \begin{subfigure}[t]{0.48\linewidth}
        \centering
        \includegraphics[
            width=\linewidth,
            keepaspectratio
        ]{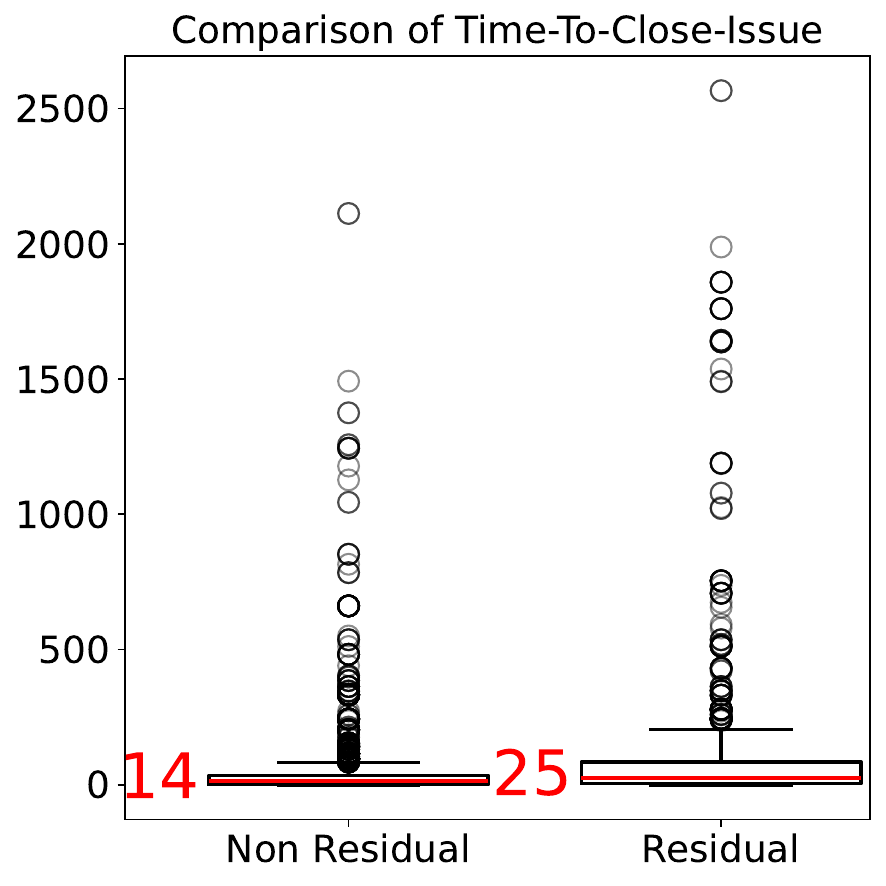}
        \caption{C/C++}
        \label{fig:bp_c}
    \end{subfigure}
    \hfill
    \begin{subfigure}[t]{0.48\linewidth}
        \centering
        \includegraphics[
            width=\linewidth,
            keepaspectratio
        ]{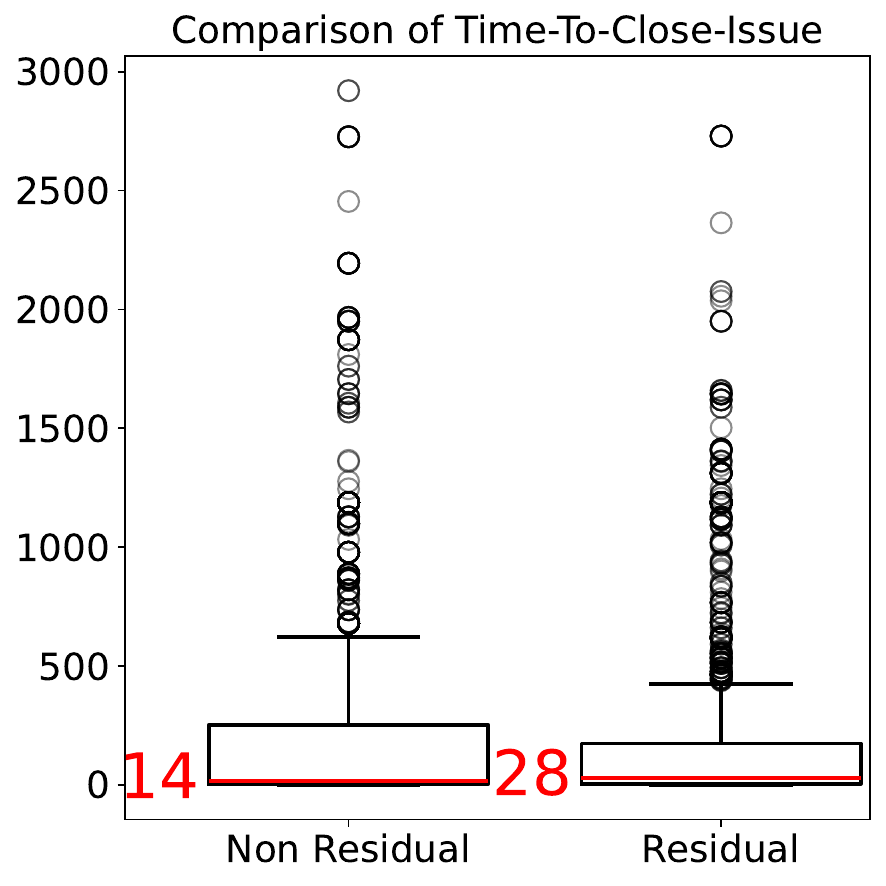}
        \caption{Java}
        \label{fig:bp_java}
    \end{subfigure}

    \vspace{0.3cm}

    \begin{subfigure}[t]{0.48\linewidth}
        \centering
        \includegraphics[
            width=\linewidth,
            keepaspectratio
        ]{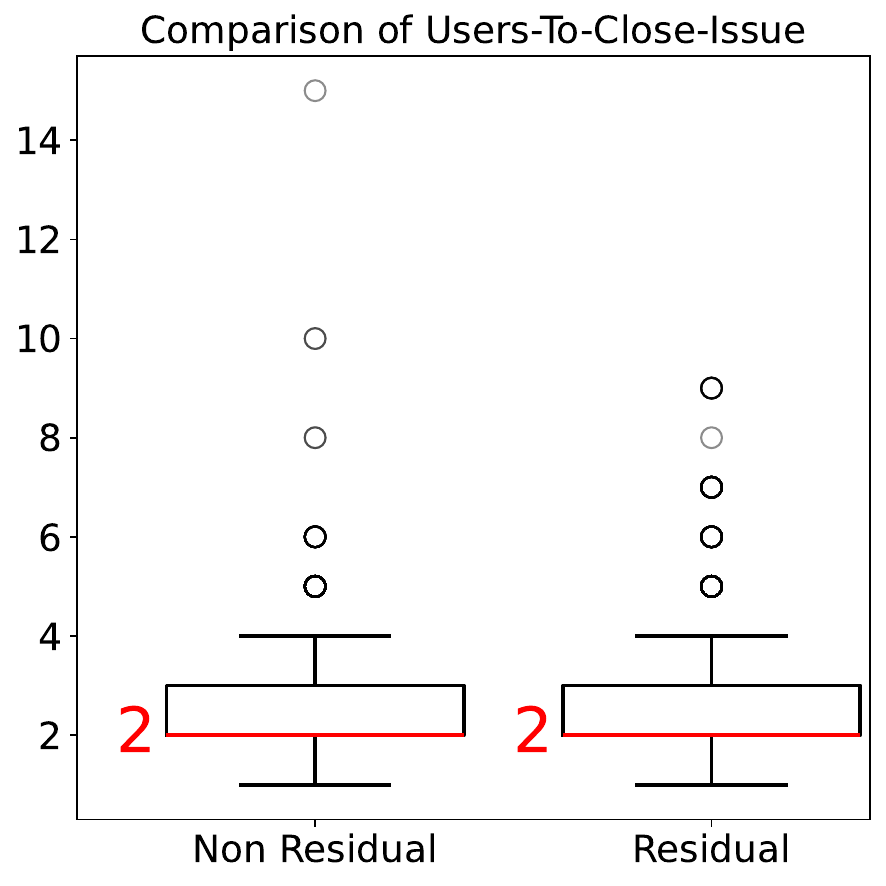}
        \caption{C/C++}
        \label{fig:bp_c_USER}
    \end{subfigure}
    \hfill
    \begin{subfigure}[t]{0.48\linewidth}
        \centering
        \includegraphics[
            width=\linewidth,
            keepaspectratio
        ]{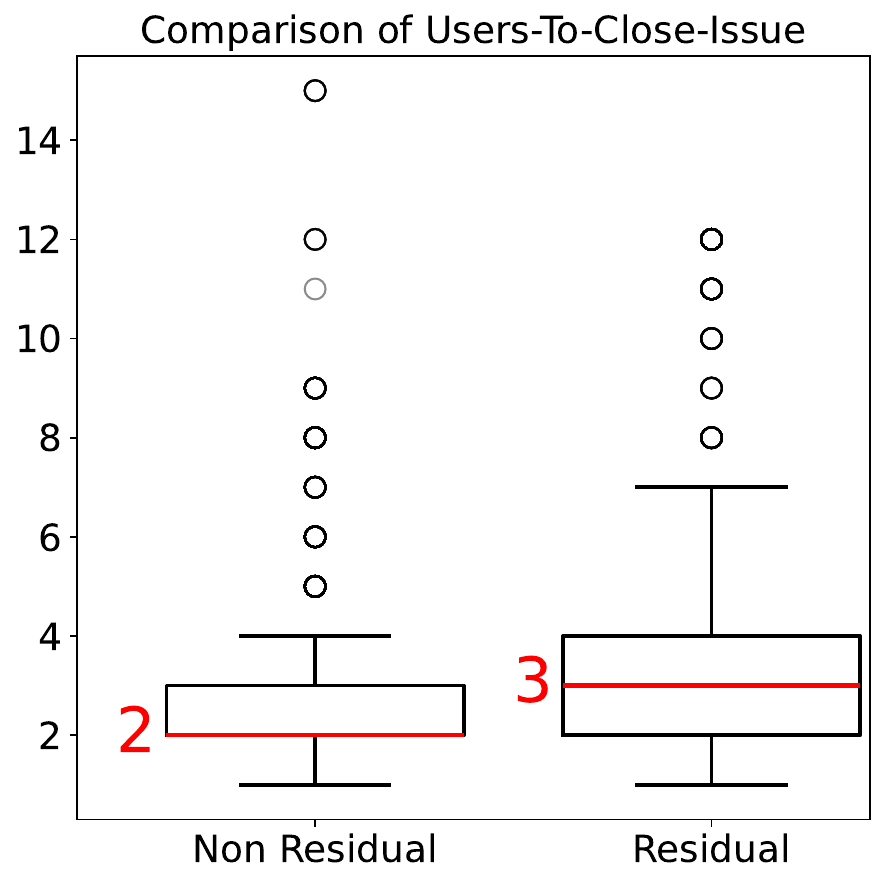}
        \caption{Java}
        \label{fig:bp_java_USER}
    \end{subfigure}

    \caption{Boxplots of \textit{Effort Metrics} in both datasets}
    \label{fig:boxplots}
    \vspace{-0.4cm}
\end{figure}

We quantify \emph{human effort} using two issue-level metrics derived from GitHub data: \textit{(i)} \textit{Time-to-Close-Issue}, calculated as the interval between issue opening and closure (expressed in days), and \textit{(ii) Users-to-Close-Issue}, measured as the number of unique contributors who commented on, closed, were assigned to, or referenced the issue. Both metrics follow standard practices in mining software repositories and were retrieved through the GitHub REST API.
We analyze the distributions of these metrics using boxplots and compare both medians (see \figurename~\ref{fig:boxplots}), as robust indicators of central tendency, and modes (as indicators of the most frequent, typical cases).

In C/C++, as shown in \figurename~\ref{fig:bp_c}, residual defects exhibit markedly higher fixing times. The median \textit{Time-to-Close-Issue} increases from 14 days for pre-release to 25 days for post-release defects, while the mode shifts dramatically from 0 to 84 days. The mode is particularly informative in this context because it reflects the most frequent cases, which is especially relevant for skewed distributions where a large portion of the data concentrates around a few recurring values.

This indicates that most issues concerning non-residual defects are closed on the same day they are opened, suggesting that many pre-release defects are relatively simple, quickly diagnosed, or already anticipated and handled as part of the regular development cycle before release.
In contrast, the most frequent closure time for post-release defects is almost three months, a substantial difference between the two groups. This shift in central tendency reinforces that post-release defects typically require more effort and time to fix.
Residual defects also exhibit a larger interquartile range, showing greater variability in closure time. This implies that C/C++ post-release defects are not only slower to fix on average but also more unpredictable, with some requiring exceptionally long diagnostic and repair efforts. Such variability reflects the heterogeneous and often environment-dependent nature of production failures in low-level systems.


In contrast, \textit{Users-to-Close-Issue} values are almost indistinguishable across groups: both have a median of 2 users and very similar interquartile ranges and upper tails (\figurename~\ref{fig:bp_c_USER}). While a few issues in each group involve many contributors, there is no clear evidence that residual defects systematically require more people to resolve; most defects, regardless of type, are handled by a small core team.

The pattern in Java differs. For \textit{Time-to-Close-Issue}, the median again increases from 14 days for non-residual to 28 days for residual defects, indicating that post-release faults typically require roughly twice as much time to diagnose, coordinate, and complete the fix compared to a non-residual one.
However, the mode remains identical across groups, offering no discriminative power: many issues in both categories share similar closure times. This suggests that the overall difference in fixing effort is smaller in Java. 

The boxplots show that the interquartile ranges of residual and non-residual defects are more distinctly separated than in C/C++, despite the modest shift in medians. This indicates that the contrast in typical effort is more consistent across Java. 


For \textit{Users-to-Close-Issue}, the median rises from 2 to 3 users in residual defects, and the upper tail extends further, suggesting a need for additional contributors. However, the overall effort remains broadly comparable across groups.

The weaker contrast observed in Java may reflect broader differences in how the two ecosystems support debugging and maintenance. Java projects typically rely on more standardized workflows, mature build and CI pipelines, and the diagnostic capabilities of the JVM, all of which tend to homogenize how defects are detected and handled. This can help explain why the most frequent resolution time (i.e., mode) is identical for pre- and post-release faults: many issues pass through similar procedural steps regardless of when they are discovered. However, the higher median closure time for residual Java defects shows that, beyond the most common cases, post-release defects still demand more effort. These issues are likely tied to less frequently exercised code paths, environment-dependent behaviors, or interactions that surface only under production workloads. Compared to C/C++, where low-level failures and memory-related faults produce a much sharper contrast, Java exhibits a subtler but still consistent separation between pre- and post-release fixing effort. 


An explanation for the weaker fixing-effort contrast in Java may lie in differences in runtime environments. C/C++ systems are often deployed in low-level or performance-critical contexts, where post-release defects involve subtle memory, concurrency, or resource interactions that are difficult to reproduce, increasing resolution effort.
Java applications, by contrast, run within managed runtimes that provide automatic memory management and stronger runtime checks, reducing certain low-level failure modes. Consequently, Java residual defects may resemble pre-release faults more closely in debugging complexity, resulting in a smaller observed effort gap.


These findings align with the results of \textit{RQ1}: residual defects are more often associated with components that exhibit higher structural and process complexity. Accordingly, the longer median closure times are consistent with residual faults tending to arise in code regions with more irregular metric profiles, which are harder to diagnose and repair.

To complement the effort-based analysis, we assessed how fixes modify code complexity. For each defect, we recomputed all product metrics on the fixed version and calculated \emph{$\Delta$ Metrics} as the difference between fixed and faulty code values. This quantifies how bug fixing changes structural and cognitive complexity. As in previous analyses, we used the Kolmogorov–Smirnov test to assess distributional differences and Cliff’s $\delta$ to evaluate their practical significance.
\tablename~\ref{tab:Delta-effects} reports all $\Delta$ metrics exhibiting at least a \textit{small} effect size.


\begin{table}[t]
\centering
\caption{KS statistics and effect sizes (Cliff’s $\delta$) for metrics showing at least a \emph{small} effect when comparing residual vs.\ non-residual defects in C/C++ and Java.}
\label{tab:Delta-effects}
\scriptsize
\begin{tabular}{lcr}
\toprule
\multicolumn{3}{c}{\textbf{C/C++ Dataset}}\\
\midrule
\textbf{Metric} & \textbf{$\delta$} & \textbf{Magnitude} \\
\midrule

Lines of Code   & $|0.1929|$ & small \\
Ratio Comment to Code & $|0.1473|$ & small \\
Halstaed Vocabulary & $|0.1651|$ & small \\
Halstaed Lenght   & $|0.2109|$ & small \\
Halstaed Volume & $|0.2269|$ & small \\
Halstaed Effort  & $|0.1712|$ & small \\

\midrule
\multicolumn{3}{c}{\textbf{Java Dataset}}\\
\midrule
\textbf{Metric} & \textbf{$\delta$} & \textbf{Magnitude} \\
\midrule
Halstaed Difficulty &  $|0.1493|$ & small \\
Halstaed Distinct Operators & $|0.1669|$ & small \\
\bottomrule
\end{tabular}
\vspace{-0.2cm}
\end{table}

In C/C++, all 23 metrics exhibit statistically significant differences between residual and non-residual defects. However, only 6 show at least a small practical effect according to Cliff’s $\delta$, while the remaining metrics display negligible effects despite very low \textit{p}-values (\textit{p}$<$ 0.05). The most pronounced differences occur in $\Delta$ \textit{LOC}, $\Delta$ \textit{Ratio Comment To Code}, and several Halstead-related metrics, all showing consistent small effects. These results indicate that, although fixes for residual and non-residual defects modify code properties in statistically detectable ways, the magnitude of these modifications is modest. Overall, the structural changes introduced during bug fixing are largely comparable across the two categories, with only limited, yet statistically significant, evidence that residual defects require more substantial code transformations.


In Java, most $\Delta$ metrics also exhibit statistically significant differences between residual and non-residual defects. However, similar to C/C++, the magnitude of these differences is modest. Out of the 23 metrics analyzed, only two, i.e., $\Delta$ \textit{Halstead Difficulty} and $\Delta$ \textit{Distinct Operators}, reach a small effect according to Cliff’s $\delta$. All remaining metrics show negligible effect sizes despite low \textit{p}-values, meaning that the distributions differ in a statistical sense but not in a way that translates into a meaningful practical divergence. The metrics that do show non-negligible differences suggest that, in Java, fixes for residual defects may introduce slightly higher cognitive complexity, as reflected by the Halstead measures. Nevertheless, the absence of stronger or more widespread effects indicates that the structural changes made during fixes are broadly similar between residual and non-residual defects.

Compared to C/C++, the contrast is weaker, reinforcing the idea that the Java ecosystem tends to produce more uniform fixes, with less variation in the extent of code modifications introduced across the two categories.
This pattern is coherent with the effort-based findings, where Java also showed only weak differences between the two categories. Taken together, these findings reinforce the idea that, within the Java ecosystem, defects tend to be handled more uniformly, both in terms of human effort and in complexity of code modifications introduced during the fix.

\begin{finding}{}{}
\textbf{Key Finding 3:} \emph{Across both languages, the analysis reveals that residual and non-residual defects differ statistically in fixing effort and code-change complexity, yet only part of these differences has a meaningful practical impact. In C/C++, post-release defects take substantially longer to close, while the number of contributors involved remains nearly constant, indicating that the additional effort is primarily \textit{technical} rather than \textit{organizational}. Code-level deltas further show small but consistent increases in structural and cognitive complexity for residual defects, particularly in size- and Halstead-related metrics. In Java, the contrast is much weaker: differences in closure time are modest, contributor involvement is almost identical, and delta metrics exhibit minimal practical variation. These findings suggest that the extra effort associated with residual defects stems mainly from technical complexity rather than expanded collaboration, with this effect being most pronounced in C/C++.} 
\end{finding}{}

\section{Discussion}
\label{sec:discussion}

Residual faults survive testing not because they are structurally different, but because their activation conditions are embedded in the evolutionary history of a component. Across both languages, age, churn, coupling, authorship dispersion, and historical bug density dominate discriminative power, whereas traditional structural metrics show limited separation. Testing typically evaluates a static snapshot of code, while residual faults depend on historically accumulated assumptions that are not explicitly encoded in test suites.

Mature components accumulate implicit invariants about state, ordering, and interactions. A change may appear locally correct and pass tests yet violate a historical or environmental assumption that manifests only in production. The longer resolution time of post-release defects, without involving more contributors, suggests that the difficulty lies in reconstructing context rather than correcting syntax. Developers do not intentionally leave such faults; rather, testing fails to recreate the evolutionary and interactional conditions that trigger them.

Entropy-based metrics provide additional insight. Although entropy correlates with general defect proneness, it does not significantly distinguish residual from non-residual faults. It captures token-level irregularity, whereas defect escape is driven by system-level and historical dynamics. This reinforces the view that residual faults are fundamentally an evolutionary reliability phenomenon.

Practically, residual defects should be treated as a distinct risk category. Testing and release processes can incorporate evolutionary risk profiling, prioritizing historically dense, high-churn, and strongly coupled components for deeper regression and integration testing. CI pipelines and release dashboards can show such signals alongside coverage metrics, making evolutionary indicators first-class reliability signals.
Defect prediction models should likewise distinguish between defect occurrence and defect escape. Because escape likelihood is more strongly associated with process and historical features than with size or complexity, models targeting production risk should emphasize churn, age and coupling.

The generalizability of these findings beyond C/C++ and Java calls for discussion. The dominance of evolutionary metrics is likely to persist across ecosystems, as churn, ownership dispersion, and historical layering are language-agnostic phenomena. However, dynamically typed languages such as Python or JavaScript may amplify interaction-related failures due to weaker compile-time guarantees and heavier reliance on runtime behavior and external dependencies. In such contexts, structural metrics may become even less discriminative, while process and dependency evolution signals grow in importance. 

Finally, AI-assisted development may amplify these patterns. LLMs often generate locally plausible code that omits corner cases or violates repository-specific assumptions~\cite{tambon2024bugs, cotroneo2025humanai, eghbali2024dehallucinator, chen2026detecting}. As AI accelerates evolution and increases churn, historically complex components may become even more prone to residual defects, strengthening the need for evolution-aware verification strategies.

\section{Threats to Validity}
\label{sec:threats}


\noindent
\textbf{Internal Validity.} 
A potential threat concerns the identification of bug-fixing commits through keyword-based heuristics, which may introduce false positives (e.g., refactorings) or false negatives (fixes without explicit keywords). Following common practice in mining software repositories, we relied on established keywords (e.g., \textit{fix, bug, patch}), which have been shown to achieve good precision in large-scale studies~\cite{sliwerski2005whendochanges, campos2017mining, yu2019characterizing}. To mitigate noise, we extract and validate function-level faulty and fixed code pairs using Tree-sitter~\cite{treesitter} parsing and remove trivial or non-behavioral edits. In addition, a multi-signal filtering stage leveraging issue labels and structured issue text patterns excludes non-defect changes.

Another threat arises from the classification of residual and non-residual defects, since issue metadata and reporter roles may be inconsistent. We mitigated this by designing a multi-signal classification heuristic that integrates issue labels, textual cues, and reporter identity, following standard practice in empirical defect classification studies. To improve precision, we explicitly discarded ambiguous cases: issues that do not expose any useful signal (e.g., missing or uninformative labels), cases where the available signals conflict, and bug–fix commits whose temporal or release information does not allow us to reliably determine whether the faulty code appears in a tagged release (e.g., inconsistent commit dates and release tags). Finally, a manual validation of a statistically representative sample of 400 issues, independently reviewed by multiple researchers, confirmed strong agreement between the heuristic and expert interpretation, thereby minimizing misclassification.

Furthermore, to minimize confounding effects due to differences in project domain or architectural complexity, all analyses comparing pre-release and post-release defects were conducted within the same projects. This design isolates defect-specific factors from project-level variation and increases the internal consistency of our comparisons.


\noindent
\textbf{Construct validity.}
The first potential issue concerns the completeness of the selected set of software metrics for this study, particularly whether additional measures (e.g., architectural or developer-related) are necessary for a comprehensive characterization of residual faults. To ensure representativeness, the metrics were drawn from a systematic review of defect prediction and fault characterization literature, covering structural, historical, developer-related, and effort-oriented dimensions validated in prior empirical work. Although higher-level architectural constructs (e.g., design smells) or fine-grained developer information (e.g., review processes) are not explicitly modeled, the selected metrics provide broad coverage while remaining automatically computable, comparable across projects, and providing strong discriminative signal. All metrics were computed using established tools and methodologies (e.g., Understand~\cite{understand}, KenLM~\cite{heafield2011kenlm}, version control data), minimizing measurement bias and ensuring consistency across Java and C/C\texttt{++} systems.

To reduce potential cross-language construct differences, identical parsing, normalization, and metric computation procedures were applied to both datasets. This ensures that observed contrasts reflect true defect characteristics rather than tool or language artifacts.

\noindent
\textbf{External Validity.}
Our datasets were drawn from open-source repositories hosted on GitHub, which may differ from proprietary or safety-critical systems in terms of development practices, testing rigor, and defect reporting. To mitigate this limitation, we selected mature and actively maintained projects with rich issue tracking histories. We also included projects from diverse domains such as libraries, frameworks, and infrastructure components to increase representativeness within open-source ecosystems. 

A second limitation is the focus on only two programming languages, C/C++ and Java. These languages represent distinct paradigms (e.g., unmanaged and managed memory), providing a useful contrast while remaining broadly representative of large-scale systems used in both open-source and industrial contexts~\cite{toplanguages}. However, the results may not directly extend to dynamically typed languages such as Python or JavaScript. We therefore interpret our findings as representative of large-scale statically typed systems and encourage future work to expand the analysis to other ecosystems.

Finally, to enhance transparency and encourage replication, we publicly released the complete data and code~\cite{RP}.

\section{Conclusion}
\label{sec:conclusion}
Residual faults that escape testing and surface in the field pose a disproportionate risk to software reliability yet remain underexplored. We characterized residual and non-residual defects at the function level across large C/C++ and Java projects, building two balanced datasets enriched with product, process, and historical metrics.

Our analysis shows that residual and non-residual defects share similar metric families but differ in how these factors interact. Residual faults display more heterogeneous relationships between structural and historical properties, with process metrics such as churn, coupling, authorship, and component age playing a stronger discriminative role than traditional structural measures, particularly in Java. These results suggest that residual defects stem from complex evolutionary dynamics rather than from code size or complexity alone.
Effort-based analyses show that post-release defects can take longer to fix but involve a similar number of contributors, indicating a primarily technical burden. 
While C/C++ and Java exhibit similar overall behavior, these patterns may not generalize across all programming languages or development ecosystems.

These findings indicate that temporal and historical signals offer crucial context for understanding and preventing defects that escape testing. Emphasizing these aspects in software quality assurance can lead to more proactive and context-aware strategies for preventing faults before they reach production.

\section*{Acknowledgment}
This work has been partially supported by the COSMIC project, U-GOV 000010–PRD-2017-S-RUSSO\_001\_001.

\IEEEtriggeratref{50} 

\bibliographystyle{IEEEtran}
\bibliography{references.bib}

\end{document}